\documentclass[prl,amssymb,twocolumn,superscriptaddress,showpacs,floatfix]{revtex4}

\usepackage{graphicx}

\newcommand{\ket}[1]{\ensuremath{\vert #1 \rangle}}

\begin{document}


\title{Photon statistics from coupled quantum dots}

\author{Brian D. Gerardot}
\affiliation{Materials Department, University of California Santa Barbara, CA 93106, Santa Barbara, USA}

\author{Stefan Strauf} \email[Corresponding author: ]{strauf@physics.ucsb.edu}
\affiliation{Materials Department, University of California Santa Barbara, CA 93106, Santa Barbara, USA}
\affiliation{Department of Physics, University of California Santa Barbara, CA 93106, Santa Barbara, USA}

\author{Michiel J.A. de Dood}
\affiliation{Department of Physics, University of California Santa Barbara, CA 93106, Santa Barbara, USA}

\author{Andrey M. Bychkov} \altaffiliation[]{University of Cambridge, Cambridge CB3 0WA, UK}
\affiliation{Department of Physics, University of California Santa Barbara, CA 93106, Santa Barbara, USA}

\author{Antonio Badolato}
\affiliation{ECE Department, University of California Santa Barbara, CA 93106, Santa Barbara, USA}

\author{Kevin Hennessy}
\affiliation{ECE Department, University of California Santa Barbara, CA 93106, Santa Barbara, USA}

\author{Evelyn L. Hu}
\affiliation{Materials Department, University of California Santa Barbara, CA 93106, Santa Barbara, USA}
\affiliation{ECE Department, University of California Santa Barbara, CA 93106, Santa Barbara, USA}

\author{Dirk Bouwmeester}
\affiliation{Department of Physics, University of California Santa
Barbara, CA 93106, Santa Barbara, USA}

\author{Pierre M. Petroff}
\affiliation{Materials Department, University of California Santa Barbara, CA 93106, Santa Barbara, USA}
\affiliation{ECE Department, University of California Santa Barbara, CA 93106, Santa Barbara, USA}

\pacs{03.67.-a 78.67.Hc 78.55.Cr }


\begin{abstract}
We present an optical study of closely-spaced self-assembled InAs/GaAs quantum dots. The energy spectrum and
correlations between photons subsequently emitted from a single pair provide not only clear evidence of coupling
between the quantum dots but also insight into the coupling mechanism. Our results are in agreement with recent
theories predicting that tunneling is largely suppressed between nonidentical quantum dots and that the interaction is
instead dominated by dipole-dipole coupling and phonon-assisted energy transfer processes.
\end{abstract}

\maketitle

Semiconductor quantum dots (QDs) are nanostructures that confine electrons and/or holes in all three dimensions.
Excitons or single electron spins in QDs are promising candidates for the storage and manipulation of both classical
and quantum bits~\cite{Loss:PRA1998}. Of particular interest for applications are semiconductor QDs in which excitons
couple to photons. They display discrete energy levels~\cite{Warburton:NAT2000}, photon
anti-bunching~\cite{Michler:NAT2000}, and Rabi oscillations~\cite{Stievater:PRL2001}. The coupling of two QDs has been
proposed as a means to generate entangled photons and to realize quantum bit gate
operations~\cite{Burkard:PRB1999,Lovett:PRB2003}. The quest for qubit operations in solid-state has triggered several
studies of the coupling processes in QD ensembles~\cite{Kagan:PRB1996} and individual QD
pairs~\cite{Schedelbeck:Science1997,Bayer:SCI2001,Ortner:PRB2005, Krenner:PRL2005}. Theoretical investigations predict
that coupling between QDs can be caused by electron and/or hole tunneling~\cite{Burkard:PRB2000,Bayer:SCI2001} or by
dipole-dipole interaction of excitons~\cite{Lovett:PRB2003,Govorov:2005,Biolatti:PRL2000,Nazir:PRB2005}. Initial
experiments reported large energy splittings up to 50 meV, attributed to tunnel coupling in identical
QDs \cite{Bayer:SCI2001}. It is now understood that the dominant effect is the different size/strain situation
of the individual QDs and that these splittings cannot be attributed to quantum mechanical coupling
\cite{Ortner:PRB2005}. Refined theories that take into account the broken symmetry of nonidentical QDs predict
repulsive forces between holes located on different dots, effectively preventing tunneling of excitons
\cite{Bester:PRB2005}. We present spectroscopic and photon-correlation measurements obtained from individual
self-assembled InAs/GaAs QD pairs that demonstrate coupling between adjacent QDs and provide
insight into the coupling mechanism and energy transfer between the QDs.\\
\indent We chose to study QD pairs with a small vertical separation of 45~\AA, where the coupling is expected to be
pronounced. The InAs QDs were grown by molecular beam epitaxy on a (100) GaAs substrate via the Stranski-Krastanow
growth mode. Strain fields above each QD form nucleation centers for QDs in a second layer, leading to vertical
QD stacking.  The s-shell transitions of the QD layers are carefully tuned to nearly identical energies during
the crystal growth~\cite{Gerardot:JCG2003,QDM:Recipe}. Micro-photoluminescence (PL) spectra were recorded using a 1.25
m spectrometer equipped with a charge coupled device. The QDs were non-resonantly excited at 780 nm.
A solid immersion lens on the sample surface was used to improve the photon collection efficiency.\\
\begin{figure}[tb!]
\includegraphics[width=84mm]{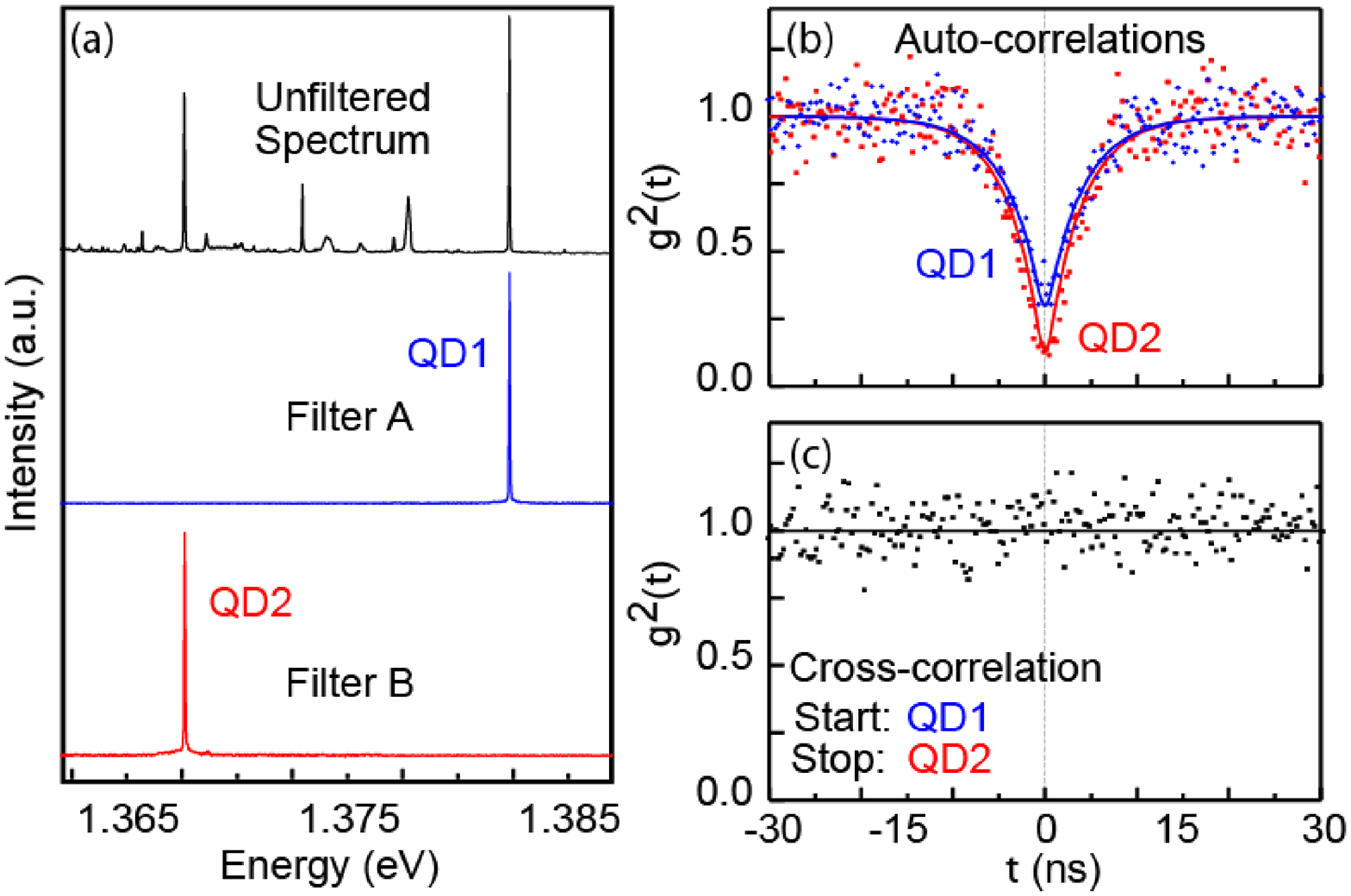}
\caption{(a) PL spectrum of a single layer of InAs/GaAs QDs. Interference filters were used to select two
emission lines from two QDs, with lateral separation of 2 $\mu$m.
(b) Photon auto-correlation for each line showing anti-bunching, $g^{(2)}(0)$~=~0.31 and 0.12 for QD1 and QD2,
respectively. (c) Corresponding cross-correlation showing $g^{(2)}(\tau)$~=~1 for all times, indicating that the
lateral QDs are uncoupled.}
\end{figure}
\indent
The experimental scheme is introduced by first describing measurements on a single layer of uncoupled InAs
QDs~\cite{KevinAntonioEvelyn}. The PL spectrum in Fig. 1a shows the emission of a few QDs at T = 4 K. The two main
emission peaks correspond to the single exciton recombination of two individual QDs laterally separated by $\sim$2
$\mu$m. To determine whether or not these two QDs are coupled the correlations between photons emitted from the QDs
have been measured. The photons pass through a fiber beamsplitter and 1 nm frequency filters in the two
output modes ($k$ and $l$) before reaching single photon detectors. Measuring the difference in arrival time between
photons at each of the two detectors provides a measurement of the second-order correlation function
$g^{(2)}_{kl}(\tau) = \langle I_k(t) I_l(t+\tau) \rangle / \langle I_k(t) \rangle \langle I_l(t) \rangle$, where
$\langle I_k(t) \rangle$ is the expectation value of the intensity at time $t$. The auto-correlation function from a
single QD is measured if both filters are tuned to the frequency of QD1 (or QD2). A single QD~\cite{Michler:NAT2000},
just like a single atom, displays photon anti-bunching at $\tau$ = 0 (Fig. 1b). For two identical but independent
two-level emitters $g^{(2)}_{k=l}(0)$ = 0.5. If the two emitters are non-identical, as is the situation here, a
post-selection of the modes $k$ and $l$ can be made using different filters to measure a cross-correlation function.
In this case of uncoupled QDs, emission of a photon from QD1
at $\tau$ = 0 does not influence a photon emission event from QD2. Therefore, the corresponding cross-correlation
function yields $g^{(2)}_{k \neq l}(\tau)$ = 1 at all delay times as confirmed by the measurement shown in Fig. 1c.
Coupling of two QDs would be characterized by measuring a deviation from 1 around $\tau$ = 0 for $g^{(2)}_{k \neq
l}(\tau)$.\\
\begin{figure}[tb!]
\includegraphics[width=84mm]{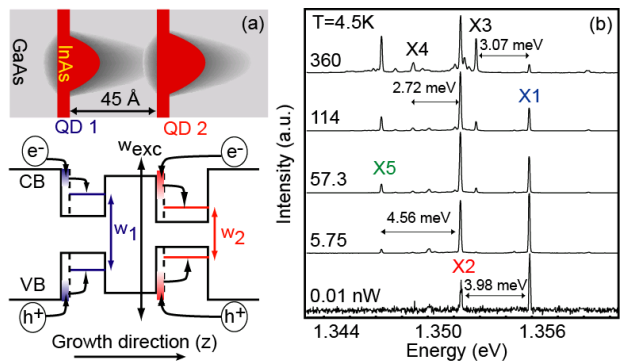}
\caption{(a) Schematic of the sample (top): two layers of InAs/GaAs QDs are vertically stacked with a separation of 45
\AA. Schematic band diagram (bottom) of the conduction band (CB) and valence band (VB) for two stacked QDs with
different transition frequencies $\omega 1$ and $\omega 2$. The pump laser, with energy $\hbar \omega_{\rm exc}$,
generates carriers in the GaAs matrix, that relax into the wetting layers and from there into the QDs. (b) Typical
power dependent PL spectra from an individual QD pair showing five dominant emission lines, labeled $X1$-$X5$.}
\end{figure}
\indent As an initial characterization, pump-power dependent spectra were taken for 20 individual pairs, all showing
similar properties. A schematic of the stacked QDs and corresponding band diagram is given in Fig. 2a. Under
non-resonant laser excitation carriers are generated in the GaAs matrix which relax into the wetting layers
and from there into the QDs. Subsequent radiative recombination leads to five dominant emission lines
in the power dependent spectra labeled by $X1$-$X5$ (Fig. 2b). At low pump powers there are two dominant peaks $X1$ and
$X2$ split by energy $\Delta E_0$ that varies from 0.5 to 5.0 meV for different QD pairs. As the pump power is
increased, the peak intensities increase linearly as expected for single exciton recombination. The spectral lines
$X3$, $X4$ and $X5$ emerge with quadratic power dependence indicative of biexcitons, while the $X1$ intensity
diminishes. The statistics on 20 individual QD-pairs with a separation of 45 \AA~demonstrates that the upper limit for
the coupling energy $\Delta E_0$ is quite small, i.e. $\leq 2.6$ meV on average. These small splitting energies are
consistent with very recent experiments using magnetic \cite{Ortner:PRB2005} and electrical field
\cite{Krenner:PRL2005} tuning, revealing anticrossing energies of ~1-2 meV. Since both coupling mechanisms, electronic
tunneling and dipole-dipole interaction \cite{Nazir:PRB2005}, will cause an energy splitting as a function of
detuning, it is not clear from such "anticrossing experiments" which mechanism will dominate the coupling.\\
\indent
To get insight into the coupling mechanism, we studied the temperature dependence of the peak intensities of
$X1$ and $X2$ transition (Fig. 3). With increasing temperature, the $X1$ intensity decreases while the $X2$ intensity
increases in such a way that the combined intensity remains constant. In addition, the measured lifetimes at
4K of the $X1$ and $X2$ states determined from auto-correlation measurements are 1.0 ns and 2.5 ns, respectively.
Both observations are indicative for a directional energy transfer from QD1 to QD2~\cite{Kagan:PRB1996,Govorov:2005}.
This directionality excludes a direct coupling between the two levels. Coupling via the continuous wetting layer states
that are $\geq$100 meV away cannot reproduce the observed strong temperature dependence. Instead, a model that couples
the $X1$ and $X2$ states through a third level,
that is $\sim$10 meV higher in energy than $X1$, can be fitted to the data (solid line in Fig. 3). The model takes into
account the decay rate of the QDs and uses temperature dependent absorption and emission rates for acoustic phonons in
thermal equilibrium to couple the levels. Self-assembled QDs are associated with extended wetting layer
states that lead to a quasi-continuous absorption background. Depending on the QD confinement potential those extended
states can approach the s-shell transition energies~\cite{Toda:PRL1999}. Evidence that these states are indeed
important in our InAs QDs have been recently reported \cite{Urbaszek:PRB2004}. Therefore, absorption of thermal energy
(acoustic phonons) can bring the exciton from QD1 into resonance with an extended state of QD2 \cite{Govorov:2005}.
From there, the excitation will quickly relax into the s-shell leading to emission of the $X2$ line and thus to a
directional energy transfer.\\
\begin{figure}[tb!]
\includegraphics[width=80mm]{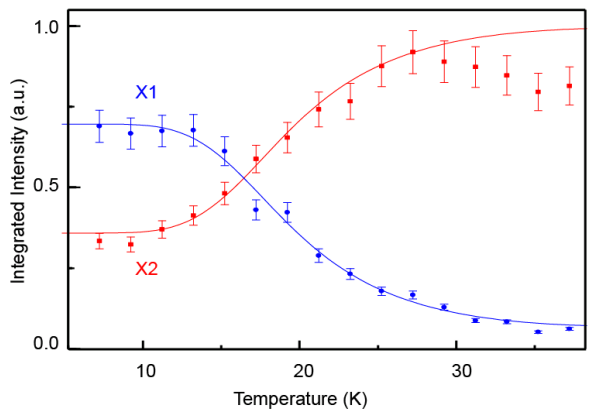}
\caption{PL for a pump power of 1.8 nW, of the $X1$ and $X2$
transitions versus temperature for a QD pair with $\Delta E_0$ = 1.3
meV. As the temperature is increased, the intensity from the
energetically higher line, $X1$, is transferred to the $X2$ line.}
\end{figure}
\indent To unambiguously demonstrate that two QDs are coupled the photon correlations between each spectral line were
studied. Auto-correlation measurements on individual spectral lines ($X1$-$X5$) all show strong anti-bunching as expected.
The main experimental results have been obtained by cross-correlation measurements between the
$X1$-$X2$, $X1$-$X5$ and $X2$-$X5$ lines, shown in Figs. 4a-c. Each cross-correlation deviates strongly from
$g^{(2)}(0)$ = 1.0, directly proving that the two QDs form a coupled system. Below a model is proposed that provides an
explanation for the observed correlations and spectral signatures.\\
\begin{figure*}[tb!]
\includegraphics[width=115mm]{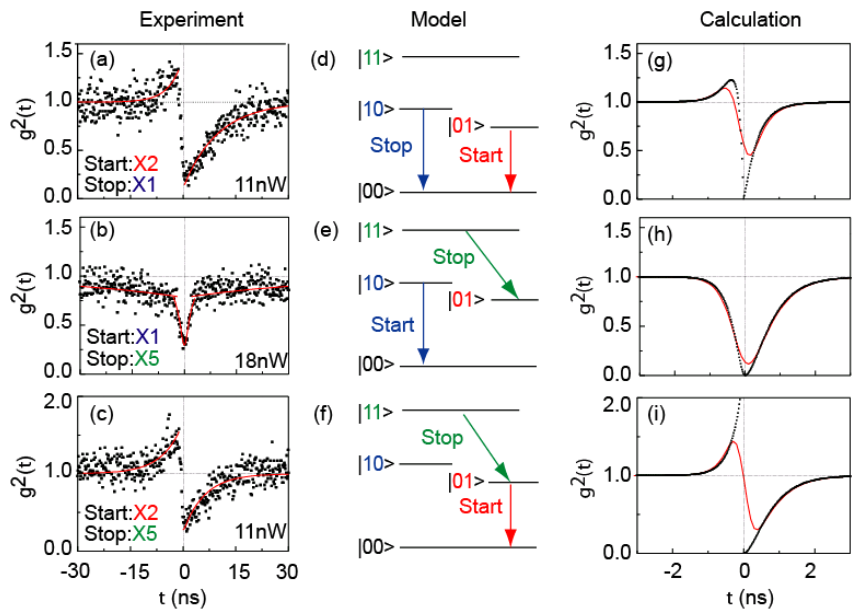}
\caption{Measured cross-correlation functions between the $X2$-$X1$, $X1$-$X5$ and $X2$-$X5$ lines (a-c) compared to
the calculated functions between the \ket{01} and \ket{10}, \ket{10} and \ket{11}, and \ket{01} and \ket{11} states
(g-i). All measurements show strong deviations from the uncoupled case $g^{(2)}(\tau)$ = 1 (see Fig. 1c), with minimum
values at zero delay time of 0.16 (a), 0.32 (b) and 0.25 (c). Figs. d-f show a level scheme of the states as used in
the model, where \ket{00} corresponds to the ground state, \ket{01} and \ket{10} to the single exciton states and
\ket{11} to the interdot biexciton state. The calculated correlation functions (dots) were convolved with a Gaussian
curve that reflects the $\sim$700 ps time resolution of the experimental system (lines). }
\end{figure*}
\indent We identify the $X1$($X2$) line as emission from the state with one exciton localized on QD1(QD2). Note that this
does include the possibility that only the hole stays localized and the electron wavefunction is spread over the double dot
structure \cite{Bester:PRB2005}.
The notation of \ket{10} and \ket{01} is introduced for these states, where the two indices denote the number of excitons present in
each QD. The states $X3$ and $X4$ emerge with quadratic power dependence at energies $\sim$3 meV less than \ket{10} and
\ket{01}, respectively. As this is the typical Coulomb binding energy for biexcitons in single InAs/GaAs QDs, these
states are labeled \ket{20} and \ket{02}. In support of this assignment, measurements (not shown) of
$g^{(2)}_{\ket{20},\ket{10}}(\tau)$ and $g^{(2)}_{\ket{02},\ket{01}}(\tau)$ exhibit the strong cascaded emission
expected for biexciton to exciton states~\cite{Moreau:PRL2001}. Finally, a new biexciton line $X5$ emerges in the
spectra at lowest photon energy. We assign this to the \ket{11} state with two excitons, one localized on each QD. This
state occurs at lowest energy due to an attractive interdot Coulomb interaction which can be larger than the
binding energy of the \ket{02} and \ket{20} intradot biexciton states.\\
\indent To model the measured photon correlations a four-level rate equation is used that includes a ground state
\ket{00}, two single exciton states \ket{10} and \ket{01}, and an inter-dot biexciton state \ket{11}. The decay rate of
the single exciton transitions $\Gamma_X$ is assumed to be equal for both dots, and directional energy transfer from
\ket{10} to \ket{01} with a rate $W_T$ is included. The pump, included via the rate $W_P$, induces transitions from the
ground state into the \ket{10} or \ket{01} states and from there into the \ket{11} biexciton state. The rate equations
for the system are:\\
\noindent $\partial \vec{n}/\partial t = \mathbf{M} \cdot \vec{n}$, where the matrix $\mathbf{M}$ is:\\
\begin{eqnarray*}
\mathbf{M} = \left(
\begin{array}{c c c c }
-2 W_P & \Gamma_X & \Gamma_X & 0 \\ W_P & -W_P - \Gamma_X & W_T & \Gamma_X \\ W_P & 0 & -W_P-\Gamma_X-W_T &
\Gamma_X
\\ 0 & W_P & W_P & -2 \Gamma_X
\\
\end{array}
\right)
\end{eqnarray*}
The column vector $\vec{n}$ ($(n_{\ket{00}},n_{\ket{01}},n_{\ket{10}},n_{\ket{11}})$) corresponds to the expectation
value to be in a particular state at time $t$. Coherences induced through the pump field are neglected as the pump
field is far off resonance. The corresponding correlation functions for the parameters $\Gamma_X$ = 2 ns, $W_P$ = 0.75
ns$^{-1}$ and $W_T$ = 5.25 ns$^{-1}$ are shown in Fig. 4g-i and qualitatively agree with the experimental measurements.
In addition, the value for the energy transfer rate is in good agreement with recent calculations assuming a phonon-assisted Coulomb transfer
for two nonidentical QDs with 4-5 nm vertical separation, 2-3 meV energy separation and a lattice temperature of 4K \cite{Govorov:2005}.
The cross-correlation results are interpreted as follows: emission of the $X1$ or $X2$ line projects the system from
the \ket{01} or \ket{10} to the ground state at $\tau$ = 0. The $X1$ to $X2$ cross-correlation in Fig. 4a is then given
by the repopulation of the \ket{01} or \ket{10} state~\cite{Berglund:PRL2002, Govorov:2005} and shows anti-bunching.
The cross-correlations in Figs. 4b and 4c involve the $X5$ emission from the \ket{11} interdot biexciton state into the
\ket{01} state. After spectral post-selection of the $X5$ emission the system is never in the \ket{10} state and the
$X1$/$X5$ (\ket{10}/\ket{11}) cross-correlation (Fig. 4b) shows pronounced anti-bunching. Conversely, cascaded emission
is observed for the $X2$/$X5$ (\ket{01}/\ket{11}) cross-correlation in Fig. 4c, similar to the quantum cascade of a
biexciton to single exciton in single QDs~\cite{Moreau:PRL2001}. Note that the model predicts in addition a transition from
the \ket{11} to the \ket{10} state, however, positive identification of this transition was hindered by the lack of intensity
required for the cross-correlation measurement.
While this model qualitatively explains the experimental results, it does not provide an explanation for the
long recovery times of $\sim$6 ns in Figs. 4a and 4c. A possible explanation is the presence of additional
meta-stable states. These can be charged or dark excitons or can be formed via tunneling of the electron only, which
is not distinguishable in the current experiment.\\
\indent
We have studied the coupling between two closely spaced self-assembled QDs with carefully tuned s-shell
transitions to nearly identical energies. The results support the prediction that exciton interdot tunneling is largely suppressed
due to the broken symmetry of nonidentical QDs \cite{Bester:PRB2005}, and that the QD coupling is instead
dominated by dipole-dipole interactions \cite{Govorov:2005}. In particular, we found a direct Coulomb interaction between
the permanent excitonic dipole moments (interdot biexciton), and a directional energy transfer between the QDs, even at
their smallest vertical separation of 45~\AA.\\
\indent
We acknowledge S. Anders, A. Imamoglu, R. Liu, L. Sham, J. Urayama, S. Yaida, and P. Zoller for fruitful discussions.
S. Strauf acknowledges support from the Max-Kade Foundation. This research has been supported by
DARPA no. MDA972-01-1-0027, NSF NIRT no. 0304678, and AFOSR no. F49620-98-1-0367 grants.


\begin{thebibliography}{30}
\bibitem{QDM:Recipe}
The InAs islands were partially covered with GaAs and annealed before complete capping to
blue-shift the QD s-shell transition energy to $\sim$920 nm. We used a 10 \AA~ and 20 \AA~GaAs
partial coverage for the first and second QD layer respectively, to tune the s-shell
transitions.
\bibitem{KevinAntonioEvelyn}
The QDs are embedded in a photonic crystal membrane that enhances the photon collection
efficiency. The QDs transitions are far from resonance with the cavity mode.
\end{thebibliography}

\end{document}